\documentclass[twocolumn,nofootinbib,amsmath,amssymb,aps,prd,balancelastpage,superscriptaddress]{revtex4-1}

\usepackage{color}
\usepackage[active]{srcltx}
\usepackage{amsmath,amsfonts,amssymb,amsthm,amstext,amscd,eucal,srcltx}
\usepackage{epsfig,graphicx,bm}
\usepackage{epstopdf, epsf}
\usepackage{dcolumn}
\usepackage{hyperref}

\newcommand{\be}{\begin{equation}}
\newcommand{\ee}{\end{equation}}

\newcommand{\bse}{\begin{subequations}}
\newcommand{\ese}{\end{subequations}}
\newcommand{\bea}{\begin{eqnarray}}
\newcommand{\eea}{\end{eqnarray}}
\newcommand{\ba}{\begin{array}}
\newcommand{\ea}{\end{array}}
\newcommand{\bc}{\begin{center}}
\newcommand{\ec}{\end{center}}

\begin{document}
\vspace*{3mm}

\title{NANOGrav results and Dark First Order Phase Transitions}%

\author{Andrea Addazi}
\email{addazi@scu.edu.cn}
\affiliation{Center for Theoretical Physics, College of Physics Science and Technology, Sichuan University, 610065 Chengdu, China}
\affiliation{INFN sezione Roma {\it Tor Vergata}, I-00133 Rome, Italy}

\author{Yi-Fu Cai}
\email{yifucai@ustc.edu.cn}
\affiliation{CAS Key Laboratory for Researches in Galaxies and Cosmology, Department of Astronomy, University of Science and Technology of China, Hefei, Anhui 230026, China}
\affiliation{School of Astronomy and Space Science, University of Science and Technology of China, Hefei, Anhui 230026, China}

\author{Qingyu Gan}
\affiliation{Center for Theoretical Physics, College of Physics Science and Technology, Sichuan University, 610065 Chengdu, China}

\author{Antonino Marciano}
\email{marciano@fudan.edu.cn}
\affiliation{Department of Physics \& Center for Field Theory and Particle Physics, Fudan University, 200433 Shanghai, China}
\affiliation{Laboratori Nazionali di Frascati INFN Via Enrico Fermi 54, Frascati (Roma), Italy}

\author{Kaiqiang Zeng}
\affiliation{Center for Theoretical Physics, College of Physics Science and Technology, Sichuan University, 610065 Chengdu, China}

\begin{abstract}
\noindent
The recent NANOGrav  evidence of a common-source stochastic background provides a hint to Gravitational Waves (GW) radiation from the Early Universe. We show that this result can be interpreted as a GW spectrum produced from First Order Phase Transitions (FOPTs) around a temperature in the KeV-MeV window. Such a class of FOPTs at temperatures much below the electroweak scale can be naturally envisaged in several Warm Dark Matter models such as Majoron Dark Matter. 

\end{abstract}

\maketitle


{\bf Introduction}. The NANOGrav Collaboration has recently released an analysis of pulsar timing data gathered over the last $12.5\,{\rm yrs}$ \cite{NANO125}. An evidence of a stochastic spectrum was found, compatibly with Gravitational Wave (GW) signals with frequency within $f\sim 10^{-6}\div 10^{-5}\, {\rm mHZ}$, average GW energy density $\langle \Omega_{GW} h^{2}\rangle_{NANOGrav}  \sim 10^{-10}$ and almost flat GW spectrum $\Omega_{GW}(f) h^{2}\sim f^{-1.5\pm 0.5}$ at $1\sigma$-level. Such a signal may be interpreted as a hint of new physics beyond the Standard Model (SM), generating GWs from Early Universe mechanisms. 
Recently, several attempts were proposed, so as to interpret the NANOGrav observations from Cosmic Strings \cite{Ellis:2020ena,Blasi:2020mfx} and Primordial Black Holes \cite{DeLuca:2020agl,Vaskonen:2020lbd}. Indeed, it is well known that in both these cases the GW spectrum is almost flat for several orders of magnitude, allowing for a possible cross-check observation in future satellite interferometers such as LISA, TAIJI and TianQin projects (around the $mHz$ GW frequency) and the future terrestrial interferometers beyond LIGO/VIRGO such as the ET project (around $10\, {\rm Hz}$ or so) \cite{Ellis:2020ena,Blasi:2020mfx,DeLuca:2020agl,Vaskonen:2020lbd}.

GW stochastic backgrounds may notoriously arise also from {\it First Order Phase Transitions} (FOPTs), in the early Universe, recently a subject of a massive investigation 
(see e.g. Refs.~\cite{H1,H2,Caprini1,Caprini2,Huang:2016odd,Addazi:2016fbj,Addazi:2017gpt,Addazi:2017oge,Addazi:2017nmg,Addazi:2018nzm,Ellis1,J,ppa,kka,HK,MY,Breitbach:2018ddu,Croon:2018kqn,Bian:2018bxr,Huang:2017rzf,Cutting:2019zws,Greljo:2019xan,Addazi:2019dqt,MW,Alanne:2019bsm,Alves:2018oct,Alves:2019igs,Wang:2020jrd}).  
The most part of the FOPTs hitherto analyzed in literature was around the electroweak (EW) scale, resulting in a test of possible SM extensions of the Higgs sector, leading to GW signals around the mHz frequencies with implications 
for future space interferometers. Within these scenarios, typically extra scalars beyond the SM are introduced as strongly coupled to the Higgs bosons, compatible with LHC constraints --- see e.g. Ref.~\cite{Addazi:2019dqt}. 

Nevertheless, FOPT dynamics can be {\it hidden} in a Dark Matter sector, weakly coupled to the Higgs and the other SM fields. In this case, FOPTs can be easily recovered at much lower energies than the EW scale, without any collider constraint. Indeed, there are several Dark Matter models, beyond the Weak Interacting Massive Particle (WIMP) paradigm, that introduce new hidden scalar fields which may have false and true minima far below the EW domain.

Alternatively to WIMPs, Dark Matter particles can have a mass much below the $10\, {\rm GeV}\div 1\, {\rm TeV}$. For example, Warm Dark Matter (WDM) particles typically have a masses in a broad range around the $0.1\, {\rm KeV}\div {\rm 100\, MeV}$. In WDM models several new extra scalars, with $0.1\, {\rm KeV}\div {\rm 100\, MeV}$ vacuum expectation values (vevs) related to new symmetries beyond the SM, can be envisaged. An explicit example is the Majoron Dark Matter model \cite{M1,M2,M3,M4,M5}, analyzed in our previous companion paper\footnote{A $0.1\div 100\, {\rm KeV}$ FOPT takes places between the Big Bang Nucleosynthesis (BBN, $T_{BBN}\sim 0.1\div 1\, {\rm MeV}$) and the Recombination ($T_{CMB}\sim 1\, {\rm eV}$) epochs. It does not alter the predictions for both the BBN and the Cosmic Microwave Background (CMB) radiation: the BBN happens much before the KeV-scale, producing the right amount of light nuclei ratios; the CMB radiation is emitted when the KeV-FOPT already ended, with no bubbles having survived and with a GW spectrum produced that is much below the CMB GW limits \cite{Addazi:2017nmg}.} \cite{Addazi:2017nmg}, for which we proposed tests in radio-astronomy experiments, including NANOGrav \footnote{ An alternative possibility of low GW frequency signals is related to Solitosynthesis of Q-balls predicted in Asymmetric Dark Matter scenarios \cite{Croon:2019rqu}.}

Indeed, if the FOPT nucleation temperature was around the WDM-scale, then the GW spectrum should be detected around the $1\div 10\, {\rm nHZ}$ frequency region, corresponding to Pulsar timing tests \cite{Addazi:2017nmg}. In principle, one would naively expect that the FOPT spectrum cannot work as an explanation of the NANOGrav excess: typically FOPT GW signals do not have a large flat plateau, as in the Cosmic Strings and Primordial Black Holes (PBH) generation cases. On the other hand, the NANOGrav sensitivity includes only an order of magnitude in the frequency window. This allows to suspect that also GW stochastic background radiation with a ``locally'' flat spectrum within the NANOGrav frequency window may provide a possible explanation for the puzzling excess. 

In this letter, we show that WDM-inspired FOPTs can explain the NANOGrav results, potentially opening a new phenomenological channel for Multi-messenger Dark Matter particle searches. We find examples of FOPTs that are compatible with the NANOGrav stochastic signal within a $95\%$ C.L. . We show that, contrary to Cosmic Strings and PBH genesis mechanisms, the FOPT spectra explaining NANOGrav will elude other GW observations at higher frequencies. Thus, in principle, a comparative analysis of FOPTs, Cosmic Strings and PBHs can allow to discriminate them from one another, thanks to future cross-correlated radio- and GW- astronomy analyses. 
 
 \vspace{0.1cm}
 
{\bf GW from FOPTs}. The GWs originated from FOPTs 
are characterized by a limited set of parameters.
The strength of the FOPT $\alpha\equiv \alpha(T)$ at the bubble nucleation temperature $T_{n}$ is related to the trace anomaly \cite{H1,H2,Caprini1,Caprini2} and casts
\begin{equation}
\label{alpha}
\alpha=\frac{1}{\rho_{\gamma}}\Big[V_{I}-V_{F}-\frac{T_{n}}{4}\Big(\frac{\partial V_{I}}{\partial T}-\frac{\partial V_{F}}{\partial T}\Big)\Big]\,,
\end{equation}
where $\rho_{\gamma}=(\pi^{2}/30)g_{*}T^{4}_{n}$ is the radiation energy density at the bubble nucleation temperature, $g_{*}$ the number of relativistic degrees of freedom, and $V_{I,F}$ are respectively the effective potentials just before and after the FOPT. Then $V_{I,F}$ do correspond, respectively, to the symmetric and broken phases. 
The characteristic rate $\beta$ of the phase transition, compared to the Hubble rate, is another important parameter impacting on the GW spectrum, expressed by 
\begin{equation}
\label{betah}
\frac{\beta}{H}=T_{n}\frac{\partial}{\partial T}\Big(\frac{S_{3}}{T}\Big)\Big|_{T_{n}}\, , 
\end{equation}
where $S_{3}$ is the thermal-corrected Euclidean action of the scalar field. 

The typical contributions to GWs from FOPTs are provided by: i) Bubble-Bubble collisions; ii) Magnetohydrodynamic (MHD) turbulence; iii) Sound shock waves (SW) in the plasma. The latter two effects are generated by the bubble violent expansion inside the Early Universe plasma. 
The three contributions are produced in a rapid transient 
of time, close to the Bubble nucleation epoch. Subsequently, they are redshifted by the Universe expansion, appearing to today observers as a 
Cosmic Gravitational Wave Stochastic Background. 

A strong and detectable GW signal can be produced if the Bubble Wall velocity is high enough. For a supersonic detonations, the velocity reads 
\begin{equation}
\label{vJ}
v_{B}=\frac{1}{1+\alpha}\Big(c_{s}+\sqrt{\alpha^{2}+\frac{2}{3}\alpha}\Big)\, , 
\end{equation}
$c_{s}=1/\sqrt{3}$ denoting the characteristic speed of sound in the plasma. Eq.~\eqref{vJ} provides a relation between the wall velocity and the FOPT strength magnitude. 

In our analysis, we will study FOPTs related to non-runnaway bubbles. It is known that in these cases, most of FOPTs predict 
a dominance of the sound waves and turbulence contributions over the collision GW spectrum \cite{H1,H2,Ellis:2018mja,Caprini2,Hoeche:2020rsg} \footnote{Recent numerical results evidence that 
the sound wave contribution has a limited transient of time 
before the MHD turbulence would induce a decoherence effect on it \cite{Guo:2020grp,Hindmarsh:2020hop,Ellis:2019oqb,Ellis:2018mja,Ellis:2020awk}. 
This leads to a revisit of GW spectral features considered in literature before \cite{Guo:2020grp,Hindmarsh:2020hop,Ellis:2019oqb,Ellis:2018mja,Ellis:2020awk}.
For further improvements on the collision contribution beyond the leading-order thin wall approximation see Refs. \cite{Lewicki:2020jiv,Ellis:2020nnr,Konstandin:2017sat}
and for progresses in lattice realisation Refs.\cite{Cutting:2020nla,Cutting:2018tjt}}.

\vspace{0.1cm}

{\bf Majoron Dark Matter}. As an example of WDM model which can generate FOPTs, we may consider a minimal extension of the SM symmetry with an extra global $U_{L}(1)$ symmetry. We introduce a new complex scalar singlet field that is a singlet of the SM gauge group and is coupled to neutrinos and the Higgs boson as 
\begin{equation}
\label{cooa}
fH\bar{L}\nu_{R}+h_{L,R} \bar{\nu}_{L,R}\nu_{L,R}^{c} +V(\sigma,H)+h.c\, , 
\end{equation}
with $h,f$ coupling matrices. The potential $V(\sigma,H)$ induces a $vev$ for the scalar field that spontaneously breaks the extra $U_{L}(1)$ once $\langle \sigma \rangle=v'$. After the symmetry breaking, LH and RH (if introduced) neutrinos acquire a Majorana mass $\mu_{L,R}=h_{L,R}v'$ \cite{M1,M2,M3}. 

It is worth to note that, while the real part of the complex scalar singlet gets a mass proportional to $v'$, the imaginary part remains perturbatively massless, as a Nambu-Goldstone boson of $U_{L}(1)$. The imaginary part field $J$ of $\sigma=\phi+iJ$ is dubbed Majoron 
and, if the symmetry is softly broken by non-perturbative quantum gravity effects, it can acquire a tiny mass \cite{M4}. Thus the Majoron can provide a candidate of either WDM \cite{M4,M5} or Cold Bose-Einstein DM \cite{Reig:2019sok}. 

The scalar sector of the model has a potential 
\begin{equation}
\label{VsigmaH}
V(\sigma,H)=V_{0}(\sigma,H)+V_{1}(\sigma)+V_{2}(\sigma,H)\, , 
\end{equation}
where 
$$V_{0}(\sigma,H)=\lambda_{s}\Big(|\sigma|^{2}-\frac{v'^{2}}{2}\Big)^{2}+\lambda_{H}\Big(|H|^{2}-\frac{v^{2}}{2}\Big)^{2}$$
\begin{equation}
\label{Vzz}
+\lambda_{sH}\Big(|\sigma|^{2}-\frac{v'^{2}}{2}\Big)\Big(|H|^{2}-\frac{v^{2}}{2} \Big)\, . 
\end{equation}

In Eq.~\eqref{VsigmaH}, the $V_{1,2}$ potentials are higher order effective operators which can efficiently catalyze 
FOPTs. \\

Resorting to a wide literature devoted to the argument, two possibilities can be considered: i) the case of five-dimensional (5-D) operators, which softly break the $U_{L}(1)$ symmetry; ii) the case of six-dimensional (6-D) operators, with 5-D terms suppressed.

The first possibility corresponds to 
\begin{equation}
\label{fi}
V_{1}^{(5)}=\frac{\lambda_{1}}{\Lambda}\sigma^{5}+\frac{\lambda_{2}}{\Lambda}\sigma^{*}\sigma^{4}+\frac{\lambda_{3}}{\Lambda}(\sigma^{*})^{2}\sigma^{3}+h.c\,,
\end{equation}
$$V_{2}^{(5)}(\sigma,H)=\frac{\beta_{1}}{\Lambda}(H^{\dagger}H)^{2}\sigma+\frac{\beta_{2}}{\Lambda}(H^{\dagger}H)\sigma^{2}\sigma^{*}$$
\begin{equation}
\label{fib}
+\frac{\beta_{3}}{\Lambda}(H^{\dagger}H)\sigma^{3}+h.c.\, ,\, 
\end{equation}
while the second scenario amounts to 
$$V_{1}^{(6)}(\sigma)=\frac{\gamma_{1}}{\Lambda^{2}}\sigma^{6}+\frac{\gamma_{2}}{\Lambda^{2}}\sigma^{*}\sigma^{5}+\frac{\gamma_{3}}{\Lambda^{2}}(\sigma^{*})^{2}\sigma^{4}$$
\begin{equation}
\label{fiia}
+\frac{\gamma_{4}}{\Lambda^{2}}(\sigma^{*})^{3}\sigma^{3}+h.c.\, , 
\end{equation}
$$V_{2}^{(6)}(\sigma,H)=\frac{\delta_{1}}{\Lambda^{2}}(H^{\dagger}H)^{2}\sigma^{2}+\frac{\delta_{2}}{\Lambda^{2}}(H^{\dagger}H)^{2}\sigma^{*}\sigma$$
$$+\frac{\delta_{3}}{\Lambda^{2}}(H^{\dagger}H)\sigma^{3}\sigma^{*}+\frac{\delta_{4}}{\Lambda^{2}}(H^{\dagger}H)(\sigma\sigma^{*})^{2}$$
\begin{equation}
\label{fiiab}
+\frac{\delta_{5}}{\Lambda^{2}}(H^{\dagger}H)\sigma^{4}+h.c.\, . 
\end{equation}

The leading order thermal corrections to the effective potential cast
\begin{equation}
\label{Maj1}
V_{\rm eff}(\sigma,T)\simeq CT^{2}(\sigma^{\dagger}\sigma)+V(\sigma,H)\, , 
\end{equation}
with
\begin{equation}
\label{equ}
C=\frac{1}{4}\Big(\frac{m_{\sigma}^{2}}{v'^{2}}+\frac{m_{H}^{2}}{v^{2}}+h^{2}_{L}+h^{2}_{R}-24K\Big)\,,
\end{equation}
in which including 5-D operators one finds
\begin{equation}
\label{corr1}
K=K^{(5)}=(\lambda_{2}+\lambda_{3})\frac{v'}{\Lambda}+\beta_{2}\frac{v'}{\Lambda}\, , 
\end{equation}
while including 6-D perators, one obtains  
\begin{equation}
\label{corr2}
K=K^{(6)}=\frac{1}{\Lambda^{2}}[(\delta_{2}+\delta_{3}+\gamma_{2}+\gamma_{3}+\gamma_{4})v'^{2}+(\delta_{2}+\delta_{3})v^{2}]\, . 
\end{equation}

When exactly L-preserving operators are taken into account,
the only terms allowed are the ones associated to the $\gamma_{4},\delta_{2},\delta_{4}$. In the following, we will consider the L-preserving case, reducing the large parameter space we have just introduced.

\vspace{0.1cm}

{\bf GW signals and NANOGrav. }
In Fig.1, we show several GW spectra from FOPTS which lie in the NANOGrav $12.5\, {\rm yrs}$ sensitivity. Comparisons with NANOGrav $11\, {\rm yrs}$, PPTA and EPTA are also displayed. The GW spectra are performed following the methodology explained in Refs.~\cite{Caprini1,Caprini2}. In particular, we explored the case of non-runaway bubbles, dominated by the sound waves and turbulence contributions 
to the GW stochastic background \cite{H1,H2,Caprini1,Caprini2}. Several FOPT spectra with nucleation temperature around the KeV-range are considered.
We show here that many possible GW signals ``enter'' within the NANOGrav $12.5$ region with an almost flat spectra, compatible with the NANOGrav $12.5$ excess within $1\sigma$ ($65\% \, {\rm C.L.}$) and a subgroup within $2\sigma$ ($95\, \%\, {\rm C.L.}$). 
It is relevant to the purpose of our discussion that the GW spectra rapidly decay for higher frequencies, rendering their effects completely elusive for GW interferometers \cite{Addazi:2017nmg}. Furthermore, WDM-FOPT spectra rapidly decay, for several orders of magnitude, thus not being affected by the constraints on the CMB radiation \cite{Addazi:2017nmg}.

\begin{figure}[ht]
\centerline{ \includegraphics [width=1.1\columnwidth]{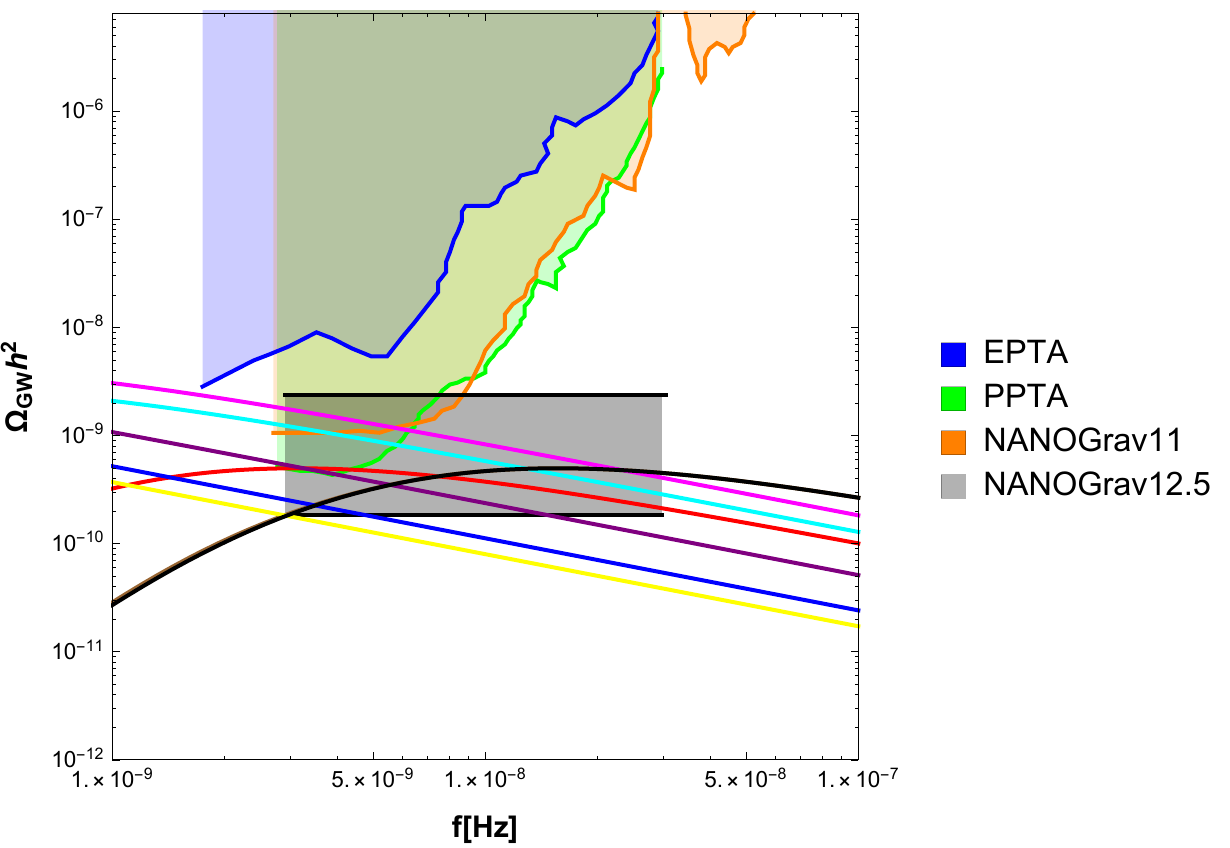}}
\caption{ Several GW signals from FOPTs are displayed and compared with the sensitivity region of NANOGrav $12.5\, {\rm yrs}$ \cite{NANO125}, NANOGrav $11\, {\rm yrs}$ \cite{NANO11}, PPTA \cite{PPTA}, EPTA \cite{EPTA}.
We show the cases of several FOPTs corresponding to different values of the $\{\alpha, \beta/H,T_{n}\}$ parameters: 1) Yellow $\{0.7,5,3\, {\rm KeV}\}$;
2) Cyan $\{0.3,10,300\, {\rm KeV}\}$; 3) Magenta $\{0.5,2,300\,{\rm KeV}\}$; 4) Blue $\{0.5,2,0.6\,{\rm KeV}\}$; 5) 
Dark purple $\{0.5,10,30\, {\rm KeV}\}$; 6) Red $\{0.1,10,2\, {\rm MeV}\}$, 7) Black $\{0.1,10,100\,{\rm MeV}\}$.  The intrinsic uncertainties of sound and turbulence efficiency factors are considered.  }
\end{figure}

The Majoron model predicts numerous FOPT points around $T_{n}=0.1{\rm KeV} \div 100 {\rm MeV}$, assuming $v'=0.1 {\rm KeV} \div 100\, {\rm MeV}$. We may also consider the effect of 6-D operators. Considering Eq.~\eqref{equ} and Eq.~\eqref{corr2}, a condition for a strong FOPT that can be detected in NANOGrav casts 
\begin{equation}
\label{bound}
K^{(6)}\simeq 4 \times 10^{-2}\Big[\frac{m_{\sigma}^{2}}{v'^{2}}+\lambda_{sH}+h_{L,R}^{2}\Big]\, . 
\end{equation}
Assuming $L$-preserving operators, only $\delta_{2},\gamma_{4}$-operators are selected, and $K^{(6)}$ simplifies to 
\begin{equation}\label{Ksim}
K^{(6)}\simeq \delta_{2}v^{2}/\Lambda^{2}\,,
\end{equation}
since the $v>\!\!>v'$. 
Assuming $\delta_{2}\simeq 1$ and $\lambda_{sH}<\!\!<1$, and having $h_{L}<\!\!<1$ for neutrino self-consistency data (while RH neutrino are integrated out at much larger energy scales), we obtain the bound on $\Lambda/v$ 
\begin{equation}\label{bound}
\Lambda \simeq 24\, v\,,
\end{equation}
namely the 6-D operator corresponding to $\delta_{2}$ has UV scale around the $4\div 5\,\, {\rm TeV}$, evading LHC limits on the electroweak sector. In our numerical analysis, performed with the help of the  {\it CosmoTransitions} package \cite{Cosmotransitions}, under the Majoron model assumptions above, we found several FOPTs within the NANOGrav 12.5 sensitivity. In particular, several FOPTs have been recovered, with $T_{n}\sim 0.1\, {\rm KeV} \div 100\, {\rm MeV} $, $\beta/H/\simeq 2\div 20$ and $\alpha\simeq 0.1\div 0.9$, that overlap the model independent plots displayed\footnote{A complete scan plot of several FOPTs allowed in the Majoron model is computationally cost-full, thus we only report some indicative branch GW solutions in Fig.~1.} in Fig.~1. 

\vspace{0.1cm}

{\bf Conclusions.}
In this letter, we showed that the strong evidence of a single-spectrum stochastic background found by NANOGrav $12.5\, {\rm yrs}$ can be explained as a GW signal from  First Order Phase Transitions (FOPTs) around the Warm Dark Matter physics scale. Indeed, reasons to suspect that FOPTs may be much below the electroweak scale are inspired by several WDM scenarios, where the typical new physics scale appears around the KeV-MeV energy window rather than in the electroweak domain. Our result inspires a different paradigm towards indirect dark matter searches beyond WIMPs. On the other hand, interestingly, radio-signals from FOPTs can be discriminated by ones from PBHs or Cosmic Strings from future multi-messenger cross-checks with GW direct detection experiments, including LISA, TAIJI, TianQin and ET. Thus, the NANOGrav result may inaugurate a new era towards a multi-messenger approach to fundamental physics
and Dark Matter.

\vspace{0.5cm}

{\bf Acknowledgements}. 
 We would like to thank Marek Lewicki and Graham White for interesting discussions and remarks on these subjects.
The work of A.A. is supported by the Talent Scientific Research Program of College of Physics, Sichuan University, Grant No.1082204112427.
Y.-F.C. is supported in part by the NSFC (Nos. 11722327, 11653002, 11961131007, 11421303), by the CAST-YESS (2016QNRC001), by the National Youth Talents Program of China, and by the Fundamental Research Funds for Central Universities.
A.M. wishes to acknowledge support by the Shanghai Municipality, through the grant No. KBH1512299, by Fudan University, through the grant No. JJH1512105, and by NSFC, through the grant No. 11875113.
All numerics were operated on the computer clusters {\it LINDA} \& {\it JUDY} in the particle cosmology group at USTC and the work station at CTP, Sichuan Univeristy.

\vspace{0.1cm}

{\bf Note added}. During the preparation of our letter, a possible explanation of NANOGrav from FOPTs appeared on arXiv \cite{FOPTN}. 
In our first analysis, we explored KeV-scale FOPTs while the authors of Ref.\cite{FOPTN} considered $0.1\div 100\, {\rm MeV}$ physics.
In the second version of our paper, we extended our analysis to the $0.1\,{\rm KeV} \div 100\, {\rm MeV}$-window which is compatible with Warm Dark Matter models.

\onecolumngrid


\twocolumngrid


\begin{thebibliography}{99}

\bibitem{NANO125}
Z. Arzoumanian {\it et al.} [astro-ph.HE/2009.04496].

\bibitem{Ellis:2020ena}
  J.~Ellis and M.~Lewicki,
  arXiv:2009.06555 [astro-ph.CO].

\bibitem{Blasi:2020mfx}
  S.~Blasi, V.~Brdar and K.~Schmitz,
  arXiv:2009.06607 [astro-ph.CO].

\bibitem{DeLuca:2020agl}
  V.~De Luca, G.~Franciolini and A.~Riotto,
  arXiv:2009.08268 [astro-ph.CO].

\bibitem{Vaskonen:2020lbd}
  V.~Vaskonen and H.~VeermÃ?e,
  arXiv:2009.07832 [astro-ph.CO].





\bibitem{H1}
M. Hindmarsh, S. J. Huber, K. Rummukainen, and D. J. Weir, 
Phys. Rev. D{\bf 92} no. 12, (2015) 123009, arXiv:1504.03291 [astro-ph.CO].
  
\bibitem{H2}  
M. Hindmarsh, S. J. Huber, K. Rummukainen, and D. J. Weir,
Phys. Rev. D{\bf 96} no. 10, (2017) 103520, arXiv:1704.05871 [astro-ph.CO].  
  
\bibitem{Caprini1}  
C. Caprini et al., 
JCAP {\bf 1604} no. 04, (2016) 001, arXiv:1512.06239 [astro-ph.CO].  

\bibitem{Caprini2}
C. Caprini et al., 
JCAP {\bf 2003} no. 03, (2020) 024, arXiv:1910.13125 [astro-ph.CO].

\bibitem{Huang:2016odd}
  F.~P.~Huang, Y.~Wan, D.~G.~Wang, Y.~F.~Cai and X.~Zhang,
  Phys.\ Rev.\ D {\bf 94} (2016) no.4,  041702
  doi:10.1103/PhysRevD.94.041702
  [arXiv:1601.01640 [hep-ph]].
  
\bibitem{Addazi:2016fbj}
  A.~Addazi,
  Mod.\ Phys.\ Lett.\ A {\bf 32} (2017) no.08,  1750049
  doi:10.1142/S0217732317500493
  [arXiv:1607.08057 [hep-ph]].
  
\bibitem{Addazi:2017gpt}
  A.~Addazi and A.~Marciano,
  Chin.\ Phys.\ C {\bf 42} (2018) no.2,  023107
  doi:10.1088/1674-1137/42/2/023107
  [arXiv:1703.03248 [hep-ph]].
  
\bibitem{Addazi:2017oge}
  A.~Addazi and A.~Marciano,
  Chin.\ Phys.\ C {\bf 42} (2018) no.2,  023105
  doi:10.1088/1674-1137/42/2/023105
  [arXiv:1705.08346 [hep-ph]].
  
\bibitem{Addazi:2017nmg}
  A.~Addazi, Y.~F.~Cai and A.~Marciano,
  Phys.\ Lett.\ B {\bf 782} (2018) 732
  doi:10.1016/j.physletb.2018.06.015
  [arXiv:1712.03798 [hep-ph]].
  
  
\bibitem{Addazi:2018nzm}
  A.~Addazi, A.~Marcianò and R.~Pasechnik,
  MDPI Physics {\bf 1} (2019) no.1,  92
  doi:10.3390/physics1010010
  [arXiv:1811.09074 [hep-ph]].
  
  \bibitem{Ellis1}
  J. Ellis, M. Lewicki, J. M. No, and V. Vaskonen,
   JCAP {\bf 1906} no. 06, (2019) 024, arXiv:1903.09642 [hep-ph].
  
  \bibitem{J}
K. Hashino, R. Jinno, M. Kakizaki, S. Kanemura,
T. Takahashi, and M. Takimoto,
  Phys. Rev. D{\bf 99} no. 7, (2019) 075011, arXiv:1809.04994 [hep-ph].
  
  \bibitem{ppa}
  A. Alves, T. Ghosh, H.-K. Guo, K. Sinha, and D. Vagie, 
  JHEP {\bf 04} (2019) 052, arXiv:1812.09333 [hep-ph].
  
  \bibitem{kka}
  G. Kurup and M. Perelstein, 
  Phys. Rev. D{\bf 96} no. 1, (2017) 015036, arXiv:1704.03381 [hep-ph].
  
  \bibitem{HK}
  K. Hashino, M. Kakizaki, S. Kanemura, P. Ko, and
T. Matsui, 
  Phys. Lett. B{\bf 766} (2017) 49?54, arXiv:1609.00297 [hep-ph].
  
  \bibitem{MY}
  S. Moretti and K. Yagyu, 
  Phys. Rev. D{\bf 91} (2015) 055022, arXiv:1501.06544 [hep-ph].
  
\bibitem{Breitbach:2018ddu}
  M.~Breitbach, J.~Kopp, E.~Madge, T.~Opferkuch and P.~Schwaller,
  JCAP {\bf 1907} (2019) 007
  doi:10.1088/1475-7516/2019/07/007
  [arXiv:1811.11175 [hep-ph]].
  
\bibitem{Croon:2018kqn}
  D.~Croon, T.~E.~Gonzalo and G.~White,
  JHEP {\bf 1902} (2019) 083
  doi:10.1007/JHEP02(2019)083
  [arXiv:1812.02747 [hep-ph]].
  
\bibitem{Bian:2018bxr}
  L.~Bian and X.~Liu,
  Phys.\ Rev.\ D {\bf 99} (2019) no.5,  055003
  doi:10.1103/PhysRevD.99.055003
  [arXiv:1811.03279 [hep-ph]].
  
\bibitem{Huang:2017rzf}
  F.~P.~Huang and J.~H.~Yu,
  Phys.\ Rev.\ D {\bf 98} (2018) no.9,  095022
  doi:10.1103/PhysRevD.98.095022
  [arXiv:1704.04201 [hep-ph]].
  
\bibitem{Cutting:2019zws}
  D.~Cutting, M.~Hindmarsh and D.~J.~Weir,
  Phys.\ Rev.\ Lett.\  {\bf 125} (2020) no.2,  021302
  doi:10.1103/PhysRevLett.125.021302
  [arXiv:1906.00480 [hep-ph]].
  
\bibitem{Greljo:2019xan}
  A.~Greljo, T.~Opferkuch and B.~A.~Stefanek,
  Phys.\ Rev.\ Lett.\  {\bf 124} (2020) no.17,  171802
  doi:10.1103/PhysRevLett.124.171802
  [arXiv:1910.02014 [hep-ph]].
  
  
\bibitem{Addazi:2019dqt}
  A.~Addazi, A.~Marciano, A.~P.~Morais, R.~Pasechnik, R.~Srivastava and J.~W.~F.~Valle,
  Phys.\ Lett.\ B {\bf 807} (2020) 135577
  doi:10.1016/j.physletb.2020.135577
  [arXiv:1909.09740 [hep-ph]].
  
  \bibitem{MW}
  A. Mazumdar and G. White, 
   Rept. Prog. Phys. {\bf 82} (2019), no. 7 076901, [1811.01948].
  
\bibitem{Alanne:2019bsm}
  T.~Alanne, T.~Hugle, M.~Platscher and K.~Schmitz,
  JHEP {\bf 2003} (2020) 004
  doi:10.1007/JHEP03(2020)004
  [arXiv:1909.11356 [hep-ph]].
  
\bibitem{Alves:2018oct}
A.~Alves, T.~Ghosh, H.~K.~Guo and K.~Sinha,
JHEP \textbf{12} (2018), 070
doi:10.1007/JHEP12(2018)070
[arXiv:1808.08974 [hep-ph]].

\bibitem{Alves:2019igs}
A.~Alves, D.~Gonçalves, T.~Ghosh, H.~K.~Guo and K.~Sinha,
JHEP \textbf{03} (2020), 053
doi:10.1007/JHEP03(2020)053
[arXiv:1909.05268 [hep-ph]].
  
  
\bibitem{Wang:2020jrd}
  X.~Wang, F.~P.~Huang and X.~Zhang,
  JCAP {\bf 2005} (2020) 045
  doi:10.1088/1475-7516/2020/05/045
  [arXiv:2003.08892 [hep-ph]].
  

 
 
  
\bibitem{Croon:2019rqu}
  D.~Croon, A.~Kusenko, A.~Mazumdar and G.~White,
  Phys.\ Rev.\ D {\bf 101} (2020) no.8,  085010
  doi:10.1103/PhysRevD.101.085010
  [arXiv:1910.09562 [hep-ph]].
  
  
\bibitem{Guo:2020grp}
  H.~K.~Guo, K.~Sinha, D.~Vagie and G.~White,
  arXiv:2007.08537 [hep-ph].
  
\bibitem{Hindmarsh:2020hop}
M.~B.~Hindmarsh, M.~Lüben, J.~Lumma and M.~Pauly,
[arXiv:2008.09136 [astro-ph.CO]].

\bibitem{Ellis:2019oqb}
J.~Ellis, M.~Lewicki, J.~M.~No and V.~Vaskonen,
JCAP \textbf{06} (2019), 024
doi:10.1088/1475-7516/2019/06/024
[arXiv:1903.09642 [hep-ph]].

\bibitem{Ellis:2018mja}
J.~Ellis, M.~Lewicki and J.~M.~No,
JCAP \textbf{04} (2019), 003
doi:10.1088/1475-7516/2019/04/003
[arXiv:1809.08242 [hep-ph]].


\bibitem{Ellis:2020awk}
J.~Ellis, M.~Lewicki and J.~M.~No,
JCAP \textbf{07} (2020), 050
doi:10.1088/1475-7516/2020/07/050
[arXiv:2003.07360 [hep-ph]].




\cite{Ellis:2018mja}

\bibitem{Hoeche:2020rsg}
S.~Höche, J.~Kozaczuk, A.~J.~Long, J.~Turner and Y.~Wang,
[arXiv:2007.10343 [hep-ph]].




\bibitem{Lewicki:2020jiv}
M.~Lewicki and V.~Vaskonen,
[arXiv:2007.04967 [astro-ph.CO]].

\bibitem{Ellis:2020nnr}
J.~Ellis, M.~Lewicki and V.~Vaskonen,
[arXiv:2007.15586 [astro-ph.CO]].

\bibitem{Konstandin:2017sat}
  T.~Konstandin,
  JCAP {\bf 1803} (2018) 047
  doi:10.1088/1475-7516/2018/03/047
  [arXiv:1712.06869 [astro-ph.CO]].
  
\bibitem{Cutting:2020nla}
  D.~Cutting, E.~G.~Escartin, M.~Hindmarsh and D.~J.~Weir,
  arXiv:2005.13537 [astro-ph.CO].
  
\bibitem{Cutting:2018tjt}
  D.~Cutting, M.~Hindmarsh and D.~J.~Weir,
  Phys.\ Rev.\ D {\bf 97} (2018) no.12,  123513
  doi:10.1103/PhysRevD.97.123513
  [arXiv:1802.05712 [astro-ph.CO]].


  
  
  
  
  
 \bibitem{M1}
 Y. Chikashige, R. N. Mohapatra and R. D. Peccei, Phys.
Lett. {\bf 98}B, 265 (1981).
 
\bibitem{M2}
 G. B. Gelmini and M. Roncadelli, Phys. Lett. {\bf 99}B, 411
(1981).
 
\bibitem{M3}
J. Schechter and J. W. F. Valle, Phys. Rev. D {\bf 25}, 774 (1982).

\bibitem{M4}
 E. K. Akhmedov, Z. G. Berezhiani, R. N. Mohapatra and G. Senjanovic, Phys. Lett. B {\bf 299}, 90 (1993) [hep-ph/9209285].
 
\bibitem{M5} V. Berezinsky and J. W. F. Valle, Phys. Lett. B {\bf 318}, 360 (1993) [hep-ph/9309214].

\bibitem{Reig:2019sok}
  M.~Reig, J.~W.~F.~Valle and M.~Yamada,
  JCAP {\bf 1909} (2019) 029
  doi:10.1088/1475-7516/2019/09/029
  [arXiv:1905.01287 [hep-ph]].

  
  \bibitem{NANO11}
  NANOGrav collaboration, The NANOGrav 1
  Astrophys. J. {\bf 859} (2018) 47 [1801.02617].
  
  \bibitem{PPTA}
  R. M. Shannon {\it et al.}, 
  Science {\bf 349} (2015) 1522 [1509.07320].
  
  \bibitem{EPTA}
  L. Lentati {\it et al.}, 
  Mon. Not. Roy. Astron. Soc. {\bf 453} (2015) 2576 [1504.03692].
  
  \bibitem{Cosmotransitions}
  C. L. Wainwright, 
  Comput. Phys. Commun. {\bf 183} (2012) 2006-2013, arXiv:1109.4189 [hep-ph].
  
  \bibitem{FOPTN}
  Y-Nakai, M. Suzuki, F. Takahashi, M.Yamada, 
arXiv:2009.09754.

\end{thebibliography}
\end{document}